\documentclass[a4paper]{llncs}
\usepackage[utf8]{inputenc}
\usepackage{mathpartir}
\usepackage{amsmath}
\usepackage{amsfonts}
\usepackage{mathpartir}
\usepackage{amssymb}

\usepackage{MnSymbol} 
\usepackage{stmaryrd}
\usepackage[disable]{todonotes}
\usepackage{xspace}
\usepackage{listings}

\newcommand{\Etwo}{\ensuremath{r}\xspace}
\newcommand{\Ematch}[2]{\Ecase~#1~\Eof~#2}
\newcommand{\Ecase}{\mathbf{case}}
\newcommand{\Eof}{\mathbf{of}}

\newcommand{\err}{\ensuremath{\mathrm{err}}\xspace}

\newcommand{\Henv}{\ensuremath{\sigma}\xspace}
\newcommand{\Hcnames}[1]{\ensuremath{\mathbf{cnames}(#1)}}
\newcommand{\Hfnames}[1]{\ensuremath{\mathbf{fnames}(#1)}}
\newcommand{\Hqnames}[1]{\ensuremath{\mathbf{qnames}(#1)}}

\newcommand{\arity}{\textrm{arity}}
\newcommand{\order}{\textrm{order}}

\newcommand{\dom}{\mathbf{dom}}

\newcommand{\Hpathto}[2]{\ensuremath{{#1}\!\selsquigarrow_{#2}}}
\newcommand{\termsTNTV}{\cT(\terminals,\nonterminals,\variables)}

\newcommand{\termsTV}{\cT(\terminals,\variables)}
\newcommand{\termsT}{\cT(\terminals)}

\newcommand{\Aalphabet}{\Sigma}

\newcommand{\Hpos}{Pos}

\newcommand{\nonterminals}{\opn{\cN}}

\newcommand{\rules}{\opn{\cR}}
\newcommand{\terminals}{\opn{\Sigma}}
\newcommand{\variables}{\opn{\cV}}

\newcommand{\basetype}{\circ}
\newcommand{\HprogToPmrs}{\mathfrak{T}}
\newcommand{\HprogTocaRTA}{\mathfrak{S}}

\newcommand{\pt}[1]{\todo[color=red]{\textbf{PT:} #1}}

\newcommand{\opn}[1]{\operatorname{#1}}

\newcommand{\Hzero}{\ensuremath{\mathrm{zero}}\xspace}
\newcommand{\Hsucc}{\ensuremath{\mathrm{succ}}\xspace}
\newcommand{\Hnil}{\ensuremath{\mathrm{nil}}\xspace}
\newcommand{\Hcons}{\ensuremath{\mathrm{cons}}\xspace}
\newcommand{\HLength}{\ensuremath{\mathrm{Length}}\xspace}
\newcommand{\HMain}{\ensuremath{\mathrm{Main}}\xspace}
\newcommand{\HAck}{\ensuremath{\mathrm{Ack}}\xspace}
\newcommand{\Hpair}{\ensuremath{\mathrm{pair}}\xspace}

\newcommand{\qqquad}{\quad\quad\quad}

\newcommand{\cA}{\mathcal{A}}
\newcommand{\cB}{\mathcal{B}}

\newcommand{\cG}{\mathcal{G}}

\newcommand{\cL}{\mathcal{L}}

\newcommand{\cN}{\mathcal{N}}

\newcommand{\cP}{\mathcal{P}}
\newcommand{\cQ}{\mathcal{Q}}
\newcommand{\cR}{\mathcal{R}}
\newcommand{\cS}{\mathcal{S}}
\newcommand{\cT}{\mathcal{T}}

\newcommand{\cV}{\mathcal{V}}

\title{Towards Tree Automata-based Success Types}
\author{Robert Jakob \and Peter Thiemann}

\institute{%
  University of Freiburg, Germany\\
  \{jakobro,thiemann\}@informatik.uni-freiburg.de
}

\begin{document}
\maketitle

\begin{abstract}
Error detection facilities for dynamic languages are often based on unit
testing.  Thus, the advantage of rapid prototyping and flexibility
must be weighed against cumbersome and time consuming test suite development.
Lindahl and Sagonas' success typings provide a means of static must-fail
detection in Erlang. Due to the constraint-based nature of the approach,
some errors involving nested tuples and recursion cannot be detected.

We propose an approach that uses an extension of model checking for
pattern-matching recursion schemes with context-aware ranked tree automata
to provide improved success typings for a constructor-based first-order
prototype language.
\end{abstract}

\section{Introduction}

Dynamic languages like JavaScript, Python, and Erlang are increasingly used
in application domains where reliability and robustness matters.
Their advantages lie in rapid prototyping and flexibility. However, no
errors are discovered until the erroneous code is actually
executed.

Thus, the main error detection facility in dynamic languages is
massive unit testing with high code coverage. As the development of unit
tests is cumbersome and time consuming, the lack of static analyses that
allow error detection prior to execution is one of the major drawbacks of
dynamic languages.

Success typings, originally presented by Lindahl and
Sagonas~\cite{LindahlSagonas2006}, provide an approach to statically
analyze dynamic languages without losing their benefits: Only
mismatches that definitely lead to a type error during execution, are
reported. This behavior is different to what a traditional type system
provides.

In such a traditional system, the typing $F: \tau_1\to\tau_2$ means that an
application of $F$ to an argument of type $\tau_1$ yields a result of type
$\tau_2$ if it terminates normally. However, some programs which do not
lead to run-time errors when executed, are rejected by the type system.  An
example is a conditional that returns values of different types in its
branches.

In contrast, a success type system guarantees that for all arguments
$v\notin\tau_1$, the function application $F(v)$ leads to a run-time error.
For an argument $v\in\tau_1$, success typing gives the same guarantees as
traditional typing: $F(v)\in\tau_2$ if it terminates normally.  Any
approach, however can only approximate the undecidable problem whether a
program contains errors.

The system of Lindahl and Sagonas is designed for Erlang and uses a
constraint-based algorithm to obtain and refine success types. Their types
are drawn from a finite lattice that encompasses atom types and union
types. One of the major goals of the original approach was the ability to
automatically generate documentation for functions from the inferred
success types. This goal requires small, readable types.

\subsection{Success typings in Erlang}

Erlang is a dynamically typed functional programming language with
commercial uses in e-commerce, telephony, and instant messaging.  Besides
the usual numeric and string types, Erlang includes an atom data type for
symbols and tuples for building data structures.

Many Erlang programming idioms rely on named tuples, that is, tuples where
the first component is an atom and the remaining components contain
associated data as in
\lstinline[language=erlang,breaklines=true]|{book,"Hamlet","Shakespeare"}|.
One can view named tuples as named constructors, where the first atom of
the named tuple gives the name of the constructor and the other elements
are arguments.  Thus, the given example corresponds to
\lstinline[language=erlang,breaklines=true]|book("Hamlet","Shakespeare")|.
Named tuples can be arbitrarily nested and dynamically created.

Lindahl and Sagonas' algorithm misses some definite errors based on nested
named tuples, as can be seen by the following example. Here is an
implementation of a list length function returning the zero constructor and
succ constructor instead of the built-in integers.
\begin{lstlisting}[language=erlang,firstline=5,lastline=6]
length([]) -> {zero};
length([_|XS]) -> {succ, length(XS)}.
\end{lstlisting} The
dialyzer infers the following success type for length: \[\mathrm{length} :
[any] \to \Hzero \cup \Hsucc(\Hzero \cup \Hsucc(\Hzero) \cup \Hsucc(any))
\] The argument part of the success type, $[any]$, describes that applying
length to a non-list argument yields an error and applying it to a list of
arbitrary content might succeed or fail. The result part describes that the
return value must be either $\Hzero$ or a nested tuple consisting of
$\Hsucc$ and $\Hzero$.  However, due to approximation, only nestings of
three levels are considered. The argument part is exact, i.e.\ there is no
argument of type $[any]$ for which length fails.

To highlight the impression, we create a check function that pattern
matches on a suitably nested hierarchy of named tuples.  This hierarchy of
tuples cannot be created by our length function. Applying our check
function to the result of length yields a definite error. However, the
success type system in Erlang cannot detect this error.
\begin{lstlisting}[language=erlang,firstline=8,lastline=10]
check({succ,{succ,{succ,{foo}}}}) -> 0.

test() -> check(length([0,0,0,0])).
\end{lstlisting}

\subsection{Our approach}
In this paper we focus on these errors and thus only consider programs
consisting of trees of constructors as values. Our approach models input
type and output type of a function with different models. A success type
for a function $f: \cA \to \cG$ is described with a context-aware ranked
tree automaton (caRTA) $\cA$ to describe the crash conditions of the
function and a pattern-matching recursion scheme (PMRS) $\cP$, which
essentially is a parametrized pattern-matching tree grammar, to describe
the output of the function. Both, automaton and grammar, can accept and
construct an infinite tree, respectively.

Our approach yields a PMRS for describing the output with the following
rules, where lists are modelled using a \Hnil and \Hcons constructor.
\begin{align*}
	&S~t\to Length~t\\
	&Length~\Hnil\to \Hzero\\
	&Length~(\Hcons~x~xs)\to \Hsucc~(Length~xs)
\end{align*}
Here, $S$ is a start symbol which takes an input which can be matched to
either \Hnil or \Hcons. We only focus on length's output and ignore the
crash conditions. This representation describes the full output length can
produce and is no approximation.

To represent the input consumption of the check function presented above, we generate a
context-aware ranked tree automaton $\cA$ which is able to capture the function's
crash behavior. In this simple case, the context-aware ranked tree
automaton degenerates to a Büchi tree automaton with a trivial acceptance
condition because there are no branches in the control-flow.
\begin{alignat*}{5}
	&\delta(q_{Check},&&\Hsucc) &&= q_{Check.\Hsucc.1}\\
	&\delta(q_{Check.\Hsucc.1},&&\Hsucc) &&= q_{Check.\Hsucc.\Hsucc.1}\\
	&\delta(q_{Check.\Hsucc.\Hsucc.1},&&\Hsucc) &&= q_{Check.\Hsucc.\Hsucc.\Hsucc.1}\\
	&\delta(q_{Check.\Hsucc.\Hsucc.\Hsucc.1},&&\mathrm{foo}) &&= \epsilon
\end{alignat*}
Here, $q_{Check}$ is the initial state of the automaton. This automaton
only accepts the tree matching the pattern of the check function.

In the example, the length function is called with the list
$[0,0,0,0]$. 
A trivial pattern-matching recursion scheme
$\cG$  with one rule
describes this input:
\[S\to \Hcons~\Hzero~(\Hcons~\Hzero~(\Hcons~\Hzero~(\Hcons~\Hzero~\Hnil)))\]

To check if a definite error occurs, we use an approach given by Ong and
Ramsay~\cite{OngRamsey2011}. This approach allows model checking of the
PMRS $\cP$ describing the output of the length function, with the tree
automaton representing the crash behavior of the check function.  For model
checking, we require an input to the length function which in our case is
$\cG$.  As expected, the model checking is not successful, because the PMRS
$\cP$ with the input $\cG$ cannot produce the tree required by the
automaton $\cA$.

\paragraph{Outline}
The rest of the paper is organized as follows. In
section~\ref{sec:preliminaries} we define pattern-matching recursion
schemes and other preliminaries. In section~\ref{sec:lang} we define a
prototype language to demonstrate our approach and in section~\ref{sec:transpmrs}
we transform a given program into a PMRS\@. 
We introduce context-aware ranked tree automata for infinite
trees, where the predecessors and sibling nodes are considered for
transitions, in section~\ref{sec:transcarta}. Additionally, we outline
a transformation from a given program to a caRTA\@.
Section~\ref{sec:mc} sketches the model checking of a PMRS and a caRTA as
an extension of model checking by type inference introduced by Kobayashi
and extended by Ong to pattern matching recursion
schemes~\cite{Kobayashi2009,OngRamsey2011}, to context-aware ranked tree
automata and pattern matching recursion schemes.
Section~\ref{sec:checking} applies this model checking approach
to our programs. Finally, we discuss related work in
section~\ref{sec:discussion}, and conclude in section~\ref{sec:conclusion}.

\section{Preliminaries}
\label{sec:preliminaries}

This section describes some preliminaries for higher-order and pattern
matching recursion schemes. For a detailed introduction please
see related work~\cite{Kobayashi2009,OngRamsey2011}.

We define a set of types
\[\tau,\sigma ::= \basetype \mid \tau \to \sigma\]
where $\basetype$ describes a tree and $\tau\to\sigma$ describes a function
from trees of the form $\tau_1$ to trees of the form $\sigma$.

We define order and arity of a type as
\begin{alignat*}{3}
	&\order(\basetype) = 0 &&\quad \arity(\basetype) = 0\\
	&\order(\tau\to\sigma) = \max\{\order(\tau) + 1, \order(\sigma)\} &&\quad \arity(\tau\to\sigma) = \arity(\sigma) + 1
\end{alignat*}
where $\order$ describes the nestedness of the arrow to the left and
$\arity$ the number of base types $\basetype$ on top-level.

A ranked alphabet $A$ is a set of symbols together with $\arity$ and
$\order$ functions defined on each symbol.

With $\terminals$ and $\nonterminals$ we fix two finite ranked alphabets,
where $f,g,h\in\terminals$ contains only first-order elements called
terminal symbols, and $F,G,H\in\nonterminals$ contains elements of
arbitrary order called non-terminals.  Additionally, we define a finite set
of variables $x,y,z\in\variables$.

Terms created from the just defined ranked alphabets are: Applicative terms
$\termsTNTV$ as expressions built from terminals, non-terminals, and
variables using application, patterns $\termsTV$ built by terminal symbols
and variables of type $\basetype$, and constructor terms $\termsT$ built
from terminals.  Applicative terms must be well-typed with respect to a
straightforward simple type system based on types $\tau$.

For a ranked alphabet $R$ we define an $R$-labelled tree $t$ as a mapping
from the string of positive integers $\{1,\dots,n\}^*$ to $\Sigma$, with
$n=\max\{\arity(s)\mid s\in\Sigma\}$, $\dom(t)$ prefix closed, and if $t(x) =
g$, then $\{i \mid x~i \in \dom(t)\} = \{1,\dots,\arity(g)\}$.

The substitution $a[x/t]$ replaces all occurrences of
the variable $x$ by $t$ in the term $a$. $\mathrm{FV}(t)$ denotes the free
variables of a term $t$.

The set of positions $Pos$ of a constructor term $t\in\termsTV$ is defined as:
\begin{itemize}
    \item If $t=x\in\variables$, then $Pos(t) = \{\epsilon\}$.
    \item If $t=c(t_1,\dots,t_n)$, then $Pos(t) = \{\epsilon\} \cup
            \bigcup_{i=1}^{n} \{ c.i.p \mid p \in Pos(t_i)\}$.
\end{itemize}
We can transform a position $p\in Pos$ to a string of the alphabet of
positive integers by simply omitting the constructor terms $c$ and
separator dots, e.g.\ $c.1.d.2$ is transformed to $12$. We use this
transformation implicitly. We define a partial order on $p,q\in Pos$ as $p
\leq q$ iff there exists $p'$ such that $p.p'=q$.  Given a term $t$ and a
variable $x$, $\Hpathto{t}{x}$ defines the set of all positions of the
occurrences of $x$ in $t$.  For a term $t$ and a position $p\in Pos(t)$, we
inductively define $t|_p$ as the subterm of $t$ at position $p$:
\begin{align*}
    t|_\epsilon := t & &
    c(t_1,\dots,t_n)|_{c.i.p} := t_i|_p
\end{align*}

A \textit{pattern matching recursion scheme}
(\textit{PMRS})~\cite{OngRamsey2011} is a parametrized grammar for infinite
trees. Formally, a PMRS $\cP=\langle \terminals, \nonterminals, \rules,
S\rangle$ consists of terminal symbols $\terminals$ and non-terminal
symbols $\nonterminals$ as described above, a finite set of rewriting
rules, which are either pattern rules or plain rules.
\begin{align*}
  F~x_1\dots x_n~p\to t & &
  F~x_1\dots x_n\to t
\end{align*}
It holds that non-terminal $F\in\nonterminals$, variables $x_i$, terminal patterns
$p\in\termsTV$, and a term $t\in\termsTNTV$ of base type.  The start symbol
$S$ is a distinguished non-terminal symbol of type $\basetype \to
\basetype$. Additionally, we require that assume that all PMRS we use are
well-typed and deterministic, and that $\llbracket t\rrbracket_P$ is
defined as the value tree of the PMRS $P$ with an argument $t$ according to
definitions in~\cite{OngRamsey2011}.  A PMRS $P$ is productive, if $\forall
t\in\termsT. \bot\notin\llbracket t \rrbracket_P$. We define $P\downarrow
F$ as the PMRS $P$ with the start symbol replaced by $F$.

A \textit{higher-order recursion scheme} (HORS) is a PMRS without pattern rules.

\section{A simple language and its success typings}
\label{sec:lang}

This section describes a first-order functional language and specifies a notion of
success typings for functions defined in this language.

Fig.~\ref{fig:syntax} defines the syntax of the language. A program is
a list of function definitions, where all names ($F,G,\dots$) are
different. Functions are unary, as $n$-ary functions can be emulated with an auxiliary $n$-ary
constructor as a wrapper. Expressions $e,r$ are function applications,
constructors ($c,d,\dots$), where each constructor has a fixed and finite
arity, and flat pattern-matching with a variable to match and patterns $P$.
The productions for $P$ define branches for pattern matching in a case
expression. The branch for constructor c specifies $n=\arity(c)$ variables
and the corresponding expression.  There is at most one branch for each
constructor in the branches for any case expression in a program.  Thus,
$P$ can be considered as mapping a constructor to a list of variables and
an expression.  Finally, values $w$ are trees of constructors $v$ or the
special error value \err.  We do not lose expressiveness by only allowing
flat pattern-matching~\cite{Augustsson1985}.  We require that variables are
bound at most once in a program and omit parentheses for nullary constructors.

\begin{figure}[t]
    \begin{minipage}[t]{0.45\textwidth}
	\begin{alignat*}{3}
		&p     &&::= f^* \\
		&f     &&::= F(x) = e \\
		&e     &&::= \Ematch{x}{P} \mid \Etwo \\
		&P     &&::= \epsilon \mid c \mapsto (x_1,\dots,x_n, e), P 
	\end{alignat*}
    \end{minipage}
    \begin{minipage}[t]{0.45\textwidth}
	\begin{alignat*}{3}
		&\Etwo &&::= F(\Etwo) \mid c(\Etwo_1,\cdots,\Etwo_n) \mid x \\
		&w     &&::= v \mid \err \\
		&v     &&::= c(v_1,\dots,v_n)
	\end{alignat*}
    \end{minipage}
	\caption{Syntax definition of our prototype language.}
	\label{fig:syntax}
\end{figure}

For the transformation of a function to an automaton and a PMRS, we
require a special syntactic form. This form ensures that: (1)
pattern-matching is only allowed on variables, (2) constructors only
contain variables, function applications, or constructors as subterms, and
(3) case expressions are not allowed within arguments to function calls.
These restrictions can be enforced by preprocessing.\footnote{Our goal is to analyze the code, not to run it. In a compiler, such a
transformation may not be advisable because it may lead to an
exponentially larger program.} Trivial crashes of the form $\Ematch{x}{[]}$ are detected during preprocessing.

The judgement $\Gamma\vdash e \Downarrow w$ defined in
fig.~\ref{fig:semantics} describes a big-step semantics for our language.
Given an environment $\Gamma$ mapping from
variables $x$ to non-error values $v$, an expression $e$ evaluates to a
value $w$. For constructor expressions, we evaluate each subexpression and
combine the results to a constructor value. If one of the expressions
evaluates to an error, the overall result of the constructor expression is
an error. A function application is evaluated by evaluating the argument to
a value and evaluating the function's body, which we get from an implicit
function store, with a new environment where only the argument variable is
bound.  If the function's argument evaluates to an error, the result of the
function application is an error, too.  Variables are evaluated by fetching
the corresponding value from the environment $\Gamma$. There are two rules
for case expressions: We fetch the constructor value of the variable to
match from the environment $\Gamma$ and then use the map corresponding to
$P$ to get the pattern variables and body $e$.  Then $e$ is evaluated with
an environment extended by the variables mapped to values defined in the
pattern and constructor argument, respectively. If there is no matching
pattern available, we return the error value.

We assume that every program has a function named Main. Thus, our initial
judgement for the program evaluation is $\emptyset\vdash \mathrm{Main}(v)
\Downarrow w$ with $v$ being an arbitrary non-error value as argument and
$w$ being the overall result. Functions may be called recursively.

\begin{figure}[t]
	
	\begin{mathpar}
		\inferrule[SCtor]{\Gamma \vdash e_i \Downarrow v_i}{\Gamma
			\vdash
			c(e_1,\dots,e_n) \Downarrow c(v_1,\dots,v_n)}
		\and
		\inferrule[SApp]{\Gamma \vdash e \Downarrow v \\ f(x) = e_b \\\\
		     x\mapsto v\vdash e_b \Downarrow w}
			 {\Gamma\vdash f(e) \Downarrow w}
		\and
		\inferrule[SVar]{\Gamma(x) = v}{\Gamma\vdash x \Downarrow v}
		\and
		\inferrule[SCase]{\Gamma(x)=c(v_1,\dots,v_n)\\ P(c)=(x_1,\dots,x_n,e)\\\\
		    \Gamma,x_1\mapsto v_1,\dots,x_n\mapsto v_n\vdash e \Downarrow w}
		 	{\Gamma\vdash \Ematch{x}{P} \Downarrow w}
		\and
		\inferrule[SCaseErr]{\Gamma(x)=c(v_1,\dots,v_n)\\ c \notin P}{\Gamma\vdash \Ematch{x}{P} \Downarrow \err}
		\and
		\inferrule[SCtorErr]{\exists i.\Gamma \vdash e_i \Downarrow \err}{\Gamma
   			\vdash
			c(e_1,\dots,e_n) \Downarrow \err}
		\and
		\inferrule[SAppErr]{\Gamma \vdash e \Downarrow \err}
			 {\Gamma\vdash f(e) \Downarrow \err}
	\end{mathpar}

	\caption{A big-step semantics for our unrestricted prototype language.}
	\label{fig:semantics}
\end{figure}

The only possibility to crash a program  is a pattern match failure in a
case expression. Thus, the inputs which definitely crash a function only
depend on case expressions in the function's body and potential recursive
calls. The argument part of a success typing can therefore be formulated as
the complement of the inputs which crash the function.

\begin{example}[Success Typing]
	In our prototype language the list length function from the introduction is defined as:
    \begin{align*}
        &\HLength(l) = \Ematch{l}{\Hnil \to \Hzero,~\Hcons(x,xs) \to \Hsucc(\HLength(l))}
    \end{align*}
	The semantics describe that the function crashes on any input that is
	not a list. The complement of this input is described as
    \[\Hnil, \Hcons(\top,\Hnil), \Hcons(\top,\Hcons(\top,\Hnil)),\dots\]
    and defines a valid argument part for a success typing of \HLength.
    Here, $\top$ corresponds to any term. As we already saw in the
    introduction, the result part of the success typing is any tree
    consisting of \Hzero and \Hsucc. A success typing for length, expressed
    as a fixpoint of a function on a set of terms, is thus:
    \[\mu X.\Hnil \cup \Hcons(\top,X) \to \mu Y.\Hzero \cup
    \Hsucc(Y)\]
\end{example}

\section{Transformation into PMRS}
\label{sec:transpmrs}

This section gives a transformation of a program $p$ into a PMRS
$\HprogToPmrs(p)$, such that both $p$ and $\HprogToPmrs(p)$ generate the
same output trees on equal input.

Similar to the transformation given in Kobayashi~\cite{Kobayashi2009}, let
$\Hcnames{p}$, $\Hfnames{p}$, be the ranked alphabets of constructor
symbols and function names defined in the program $p$, respectively.  We
define a transformation $\HprogToPmrs(p) = \langle \Aalphabet, \cN, \cR,
\cS\rangle$ from programs to PMRS's where $\Aalphabet = \Hcnames{p}$, $\cN
= \Hfnames{p}$, and $\cS = \HMain$. To get the rules for the PMRS, we
transform each function separately $\cR=\bigcup\{\HprogToPmrs_{F,x}(e) \mid
F(x)=e \in p\}$ where $\HprogToPmrs_{F,a}(e)$ is an auxiliary function from
function name $F$, context $a\in\termsTV$ and expression $e$ to a set of
rules:
\begin{alignat*}{4}
  &\HprogToPmrs_{F,a}(G(e)) &&= \{F~a\to G~e\} &\\
  &\HprogToPmrs_{F,a}(c(\overline{e})) &&= \{F~a \to c~\overline{e}\} &\\
  &\HprogToPmrs_{F,a}(x) &&= \{F~a\to x\} &\\
  &\HprogToPmrs_{F,a}(\Ematch{x}{c_i(\overline{y})\to e_i}) &&= \bigcup_i \HprogToPmrs_{F,a[x/c_i(\overline{y})]}(e_i) &
\end{alignat*}

In PMRS's, all symbols from $\Aalphabet$ and $\cN$ are typed. We do not
distinguish between different classes of constructors in our PMRS and thus
use $\basetype$ as single base type. The type for function symbols
$F\in\cN$ is $F : \basetype\to \basetype$, as all functions are unary.  For
constructor symbols $c_i\in\Aalphabet$ we have $c_i:
\underbrace{{\basetype}\to\cdots\to{\basetype}}_{n}$ where $n$ is the arity of the constructor
defined in our language.

\begin{example}[Transformation of a program to a PMRS]
	As an example, we transform the Ackermann function into a PMRS\@.
    Let $p$ be the program defined in fig.~\ref{fig:ack}.
    \begin{figure}[t]
	\begin{alignat*}{3}
		&\HMain(a) = &&\HAck(a)\\
		&\HAck(a) = &&\Ecase~a~\Eof \\
		&	&&\Hpair~m~n \to \Ecase~m~\Eof \\
		&	&& \qqquad\Hzero \to \Hsucc(n) \\
		&	&& \qqquad\Hsucc~x \to \Ecase~n~\Eof \\
		&	&& \qqquad\qqquad\Hzero \to \HAck(\Hpair(x,\Hsucc(\Hzero))) \\
		&	&& \qqquad\qqquad\Hsucc~y \to \HAck(\Hpair(x,\HAck(\Hpair(\Hsucc(x),y))))
	\end{alignat*}
    \caption{Definition of the Ackermann function.}
    \label{fig:ack}
\end{figure}
	Using the transformation function $\HprogToPmrs(p)$ we get
    $\langle
	\Aalphabet, \cN, \cR, \cS \rangle$ with
	\begin{align*}
		&\Aalphabet=\{\Hpair: \basetype\to\basetype\to\basetype, \Hsucc:\basetype\to\basetype, \Hzero: \basetype\}\\
		&\cN=\{\HMain: \basetype\to\basetype\} \\
		&\cS=\HMain
	\end{align*}
	and $\cR$ as:
	\begin{align*}
		&\HMain~a = \HAck~a\\
		&\HAck~(\Hpair~\Hzero~n) = \Hsucc~n\\
		&\HAck~(\Hpair~(\Hsucc~x)~\Hzero) = \HAck~(\Hpair~x~(\Hsucc~\Hzero))\\
		&\HAck~(\Hpair~(\Hsucc~x)~(\Hsucc y)) = \HAck~(\Hpair~x~(\HAck~(\Hpair~(\Hsucc~x)~y)))
	\end{align*}
	\label{ex:ackermannpmrs}
\end{example}

For the later defined model-checking approach, we require all PMRS's to be
productive. We approximate productivity of a PMRS $\cP$ using a standard
flow analysis. From now on we assume, that all PMRS's are productive.

\section{Transformation into caRTA}
\label{sec:transcarta}

This section sketches a transformation of a program into a tree automaton
to capture its crash conditions. If the resulting automaton rejects a tree,
representing a function's input, then applying the function to this tree
yields an error.

Our automata model is a context-aware ranked tree automaton (caRTA), which
is an extension of a Büchi tree automaton (BTA). BTA and caRTA differ in
their transition function. In a BTA the transition function $\delta_{\cB}:
Q\times\Sigma\to Q^*$ describes how to proceed when given a node with a
state and a symbol.  In a caRTA however, a transition may depend on a
finite context of the current node. The context may include ancestors,
siblings and their descendants up to a maximum predefined size.

\begin{definition}[Context-aware Ranked Tree Automaton]
A context-aware ranked tree automaton (caRTA) $\cA=\langle \Sigma, Q,
\delta, q\rangle$ is defined by a finite ranked alphabet $\Sigma$ defining
the input symbols, a finite set of states $Q$, a transition function
$\delta: Q\times \cT(\Sigma)\times \Hpos \to Q^*$ and an initial state $q$.
The transition function maps the current state, a context term, and a
path to the current symbol, to a set of states for the children of
the current node in the tree. We restrict our transitions such that
$\delta(q,t,p)=q_1\cdots q_n \Rightarrow t|_p=c(x_1,\dots,x_n)$ where
$n=\arity(c)$.  Furthermore, we do not allow conflicting transitions. Two
transitions $\delta(q,t,p)=\overline{q}$ and
$\delta(q,t',p')=\overline{q}'$ are conflicting, if $t$ is unifiable with
$t'|_{p''}$, where $p'=p''.p$. We define $\cA\downarrow
q$ as the caRTA $\cA$ with the initial state replaced by $q$.
\end{definition}

A $\Sigma$-labelled tree $t$ is accepted by a caRTA $\cA$ if there exists a
$Q$-labelled tree $r$ such that both trees have the same domain $\dom(t) =
\dom(r)$, and for every $x\in\dom(r)$ there is $\delta(r(x), s, p) =
r(x~1)\dots r(x~m)$ with $m=\arity(t(x))$ and for all $p'\leq p$, $t(x')$ is
unifiable with $s|_p'$ and $x=x'p'$. Additionally, the special symbol \@?
is always accepted by the automaton independent of the current state.

\begin{example}
	We define an automaton $\cA=\langle \Sigma, Q, \delta, q_0\rangle$ with
	\begin{align*}
		& \Sigma = \{a: \circ\to\circ\to\circ, b:\circ,c:\circ\} \\
		& Q = \{q_0,q_1,q_2\} \\
        & \delta(q_0, \underbar{a}, \epsilon) = q_1~q_2 && \delta(q_*,\underbar{*},\epsilon)=\overline{q_*} \\
		& \delta(q_1, a~\underbar{b}~x, a.1) = \epsilon && \delta(q_2, a~b~\underbar{*}, a.2) = \overline{q_*} \\
		& \delta(q_1, a~\underbar{c}~x, a.1) = \epsilon && \delta(q_2, a~c~\underbar{c}, a.2) = \epsilon 
	\end{align*}
    With $*$ we represent that any term may occur at this position.
    The state $q_*$ is a drain state that accepts any term
    The current node of the transition which corresponds to the path is emphasized by
    underlining.  The automaton accepts the following trees:
	\[\forall t\in\termsT.a~b~t\ \textnormal{and}\ a~c~c \]
\end{example}

The transformation from a program $p$ to a caRTA $\cA$ proceeds in three
steps: At first, we translate the different pattern-matching cases into
transitions of an automaton.  Then, we analyze the function calls
and the variable bindings and extract constraints for the variables.
Finally, we transform the automaton according to the constraints.

\subsection{First step: Creation of the automaton} As the only possibility
to crash a program is a pattern match failure, the first step only
transforms the nested pattern-matching structure of the given program and
its functions into a caRTA\@. The contexts of the caRTA are used to map the
control-flow of the program into the automaton.

We use a transformation $\HprogTocaRTA(p)=\langle \Aalphabet, Q, \delta, q
\rangle$ from a program $p$ to a caRTA, where $\Aalphabet = \Hcnames{p}$,
$Q=\Hqnames{\delta}$ which extracts all states from the transitions in
$\delta$, \pt{ok?} and $q=q_{\HMain}$. We define $\delta = \bigcup
\{\HprogTocaRTA_{\{x\},x\mapsto q_F,F,x}(e) \mid F(x)=e \in p\}$, which
calls the auxiliary function $\HprogTocaRTA_{V,\sigma,F,a}(e)$ defined in
fig.~\ref{fig:progtocarta} to transform pattern-matching cases into
transitions. The auxiliary function $\HprogTocaRTA_{V,\sigma,F,a}(e)$ maps
from a set of variables $V$, which contains the variables that have not
been pattern-matched so far, an environment $\sigma$ as mapping from
variable to state, a function symbol $F$, the current context
$a\in\termsTV$ and the current expression $e$ to a set of transitions.  For
each pattern in a case expression, a transition rule is created using the
current context. The auxiliary function is applied on every pattern body
with adjusted arguments and the resulting transitions are collected.  The
set of the not-pattern matched variable set is adjusted, and the context is
adapted because we gain information about variables through pattern
matching.  Additionally, the environment $\sigma$ is extended with
variables defined in the patterns.  The states are created using the paths
to the variables in the new context.  When an expression $r$ is
encountered, no more case expressions can occur.  Thus, we create
transitions into drain states for all variables in $V$. 

We remember the environment $\sigma_p$ which maps all
variable in $p$ to their corresponding state. The context $a$ is necessary,
because the automaton has to respect control flow in the program.

\begin{figure}[t]

    \begin{minipage}[t]{0.45\textwidth}
\begin{align*}
    &\HprogTocaRTA_{V,\sigma,F,a}(r)=\bigcup_{v\in V} \{(\sigma(v),a',pos) \to \overline{q_*}\} \qquad \\
    &\textnormal{where:}\\
    &\quad a' = a[\!*\!]_{\Hpathto{a}{v}}\\
    &\quad pos = \Hpathto{a}{v}
\end{align*}
\end{minipage}
    \begin{minipage}[t]{0.45\textwidth}
\begin{align*}
    &\HprogTocaRTA_{V,\sigma,F,a}(\Ematch{x}{c_i(\overline{x_{i}}) \rightarrow e_i})
            = \bigcup_i \delta_i \cup \delta'\\
    &\textnormal{where:}\\
    &\quad \HprogTocaRTA_{V_i,(\Henv,x_{i1}\mapsto q_{p_{i1}},\dots,x_{im}\mapsto
            q_{p_{im}}),F,a_i}(e_i)=\delta_i\\
    &\quad a_i = a[x/c_i(\overline{x_i})] \\
    &\quad V_i = V\setminus x \cup \{\overline{x_i}\} \\
    &\quad p_{ij} = \Hpathto{a_i}{x_{ij}} \\
    &\quad \delta'= \bigcup_i \{(\sigma(x),a_i, \Hpathto{a}{x})\rightarrow q_{p_{i1}}\dots q_{p_{im}}\}
\end{align*}
\end{minipage}
\caption{Transformation from an expression to a set of caRTA transitions.}
  \label{fig:progtocarta}
\end{figure}

\begin{example}[Creation of automaton]
	We apply $\HprogTocaRTA(p)$ to the Ackermann function defined in
	example~\ref{ex:ackermannpmrs} and obtain $\HprogTocaRTA(p)=\langle
	\Sigma, Q, \delta, q_{\HMain} \rangle$ as defined in
    fig.~\ref{fig:ackcarta}.

    \begin{figure}[t]
        \begin{align*}
            &\Aalphabet=\{\Hpair: \basetype\to\basetype\to\basetype, \Hsucc:\basetype\to\basetype, \Hzero: \basetype\}\\
            &Q=\{q_{\HMain}, q_{Ack}, q_{Ack.pair.1}, q_{Ack.pair.2}, q_{Ack.pair.1.succ.1}, q_{Ack.pair.2.succ.1}\} \\
            &\delta(q_{\HMain}, \underbar{*}, \epsilon) = \overline{q_*} \\
            &\delta(q_{Ack}, \underbar{\Hpair~m~n}, \epsilon) = q_{Ack.pair.1}~q_{Ack.pair.2} \\
            &\delta(q_{Ack.pair.1}, \Hpair~\underline{\Hzero}~n, \Hpair.1) = \epsilon \\
            &\delta(q_{Ack.pair.2}, \Hpair~\Hzero~\underline{*}, \Hpair.2) = \overline{q_*} \\
            &\delta(q_{Ack.pair.1}, \Hpair~(\underline{\Hsucc~x})~n, \Hpair.1) = q_{Ack.pair.1.succ.1} \\
            &\delta(q_{Ack.pair.1.succ.1}, \Hpair~(\Hsucc~\underline{*})~n, \Hpair.1.\Hsucc.1) = \overline{q_*} \\
            &\delta(q_{Ack.pair.2}, \Hpair~(\Hsucc~x)~\underline{\Hzero}, \Hpair.2) = \epsilon \\
            &\delta(q_{Ack.pair.2}, \Hpair~(\Hsucc~x)~(\underline{\Hsucc~y}), \Hpair.2) =q_{Ack.pair.2.succ.1} \\
            &\delta(q_{Ack.pair.2.succ.1}, \Hpair~(\Hsucc~x)~(\Hsucc~\underline{*}), \Hpair.2.\Hsucc.1) = \overline{q_*} 
        \end{align*}
        \caption{The caRTA for the Ackermann function after the first step.}
        \label{fig:ackcarta}
    \end{figure}

    Function calls are not considered in this step, thus for all variables
    which are not pattern-matched, drain transitions are created.
\end{example}

\subsection{Second step: Analyzing the calls} The automaton we just created
has to be adapted in a second step as it does not respect constraints
induced by function calls.  Each variable is represented as a state in our
automaton according to the mapping $\sigma_p$. Thus, we analyze use of
variables in each function call to detect equivalent states.

We define $\cQ:Q\times Q$ to be the smallest relation closed under
reflexivity, symmetry, and transitivity with:
\[\cQ = \{(\sigma_p(x),q_{F.s})\} \mid F(e) \in p, x\in FV(e), s\in\Hpathto{e}{x}\}\]
Thus, $\cQ$ is an equivalence relation created from all variables used as
arguments to function applications in the program. The current state of the
variable is obtained from the previously defined mapping $\sigma_p$.  The
new state the variable should be mapped to, is built by extracting the path
to the variable in the argument, prepended by the name of the function that
is called.

\begin{example}[Call analysis]
	The Ackermann function has three function calls where four variables
    are used. The equivalence relation is given in fig.~\ref{fig:equivacker}.

    \begin{figure}[t]
	\begin{align*}
		\cQ = \{&(q_{Ack.\Hpair.1.succ.1},q_{Ack.\Hpair.1.succ.1}), \\
			&(q_{Ack.\Hpair.1.\Hsucc.1},q_{Ack.\Hpair.1}), \\
            &(q_{Ack.\Hpair.2.\Hsucc.1},q_{Ack.\Hpair.2})\}
        \end{align*}
    \caption{Non-closed relation $\cQ$ on states for the Ackermann function.}
    \label{fig:equivacker}
\end{figure}
\end{example}

\subsection{Third step: Intersection of the automaton}
Finally, we have to adapt the automaton created in the first step to the
state equivalence relation created in the second step.

This last step is surprisingly difficult because of the necessity to retain
the contexts such that the original control-flow is unharmed. We are
working on a formalization for this step.

\begin{example}[Final automaton]
	For the Ackermann function, a possible final automaton is defined in
	fig.~\ref{fig:ackfinal}.
    \begin{figure}[t]
        \begin{align*}
            &\delta(q_{Ack}, \underbar{\Hpair~m~n}, \epsilon) = q_{Ack.pair.1}~q_{Ack.pair.2} \\
            &\delta(q_{Ack.pair.1}, \Hpair~\underline{\Hzero}~n, \Hpair.1) = \epsilon \\
            &\delta(q_{Ack.pair.2}, \Hpair~\Hzero~\underline{*}, \Hpair.2) = \overline{q_*} \\
            &\delta(q_{Ack.pair.1}, \Hpair~(\underline{\Hsucc~x})~n, \Hpair.1) = q_{Ack.pair.1'} \\
            &\delta(q_{Ack.pair.2}, \Hpair~(\Hsucc~x)~\underline{\Hzero}, \Hpair.2) = \epsilon \\
            &\delta(q_{Ack.pair.2}, \Hpair~(\Hsucc~x)~(\underline{\Hsucc~y}), \Hpair.2) =q_{Ack.pair.2'} \\
            &\delta(q_{Ack.pair.1'}, \Hzero, \epsilon) = \epsilon \\
            &\delta(q_{Ack.pair.1'}, \Hsucc~x, \epsilon) = q_{Ack.pair.1'} \\
            &\delta(q_{Ack.pair.2'}, \Hzero, \epsilon) = \epsilon \\
            &\delta(q_{Ack.pair.2'}, \Hsucc~x, \epsilon) = q_{Ack.pair.2'}
        \end{align*}
        \caption{The final automaton capturing the crash conditions for the
           Ackermann function.}
        \label{fig:ackfinal}
    \end{figure}
	This automaton can be minimized, as the transitions for
	$q_{Ack.pair.1'}$ and $q_{Ack.pair.2'}$ are equivalent.
\end{example}

\section{Model checking}
\label{sec:mc}

This section discusses the question whether the output trees from a
PMRS $\cP$ with some input defined by a HORS $G$ are accepted by a caRTA
$\cA$.

As we do not know the input to our functions in general, we use a trivial
input HORS $\cG_?$ defined as
\begin{align*}
    &\cG_? = \langle \Sigma=\{?\},\cN=\{S\},R,S \rangle\\
    &\textnormal{with}\ R: S \to ?
\end{align*}
As defined above, \@? is a special symbol always accepted by an arbitrary
caRTA.

Given a deterministic PMRS $\cP$, the HORS $\cG_?$ just defined, and a
context-aware ranked tree automaton $\cA$, we define:
\[\vDash (\cP,\cA)\quad \textnormal{iif}\quad \forall t\in \cL(\cG_?).\ \llbracket Main~t \rrbracket_\cP \in
\cL(\cA)\]
The \textit{caRTA-extended PMRS verification problem} is to decide the
truth of $\vDash(\cP,\cA)$.

Our problem is similar to the PMRS verification problem described by
Ong and Ramsay~\cite{OngRamsey2011}. We can follow their approach based on
a counter-example guided abstraction refinement (CEGAR) until the last
step: At first, an over-approximation of the given PMRS is calculated.
Then, they eliminate non-determinism by adding a family of terminal
symbols, symbolizing the non-deterministic choice and adopt the automaton.
This adoption is possible on our caRTA, too. Finally, they end up with
recursion schemes with weak-definition-by-cases (wRSC), which are equi-expressive
to the over-approximated PMRS.

The model checking of the wRSC with the automaton $\cA$ is based on a type
inference. We have to extend this type inference because we have to cope
with the contexts present in our context-aware ranked tree automaton.

\subsection{Model checking by type inference}
Model checking by type inference~\cite{OngRamsey2011} is based on an
intersection type system. 

We plan on extending this type system as follows: Let $n$ be the maximum
height of the contexts used in $\cA$.  We required our initial PMRS to be
productive, thus, wRSC is productive, too. Then, from the wRSC $\cG$, we
approximate the set of trees of height $n$ that occur in the value tree of
$\cG$ as $T$. Finally, we have to adapt the \textsc{Term} rule from the
intersection type system such that it checks whether the contexts requested
from the automaton occur in $T$.

\section{Checking programs for definite errors}
\label{sec:checking}

This section sketches an idea how to find definite errors in a given
program $p$ using the model checking possibilities described above.

From our program $p$ we create a PMRS $\HprogToPmrs(p)$ and a caRTA
$\HprogTocaRTA(p)$. For every function application $F(t)$ we check for
definite errors
\begin{itemize}
	\item If the argument contains no function application, we check if
		if $\HprogTocaRTA(p)\downarrow q_F$ accepts by the argument $t$
		where all variables are replaced by \@?.
	\item If the argument contains $G(t')$ with a possible prefix $c$, we
		insert a new rule $F' \_ \to t$ and verify that
		$\vDash (\HprogToPmrs(p)\downarrow G,\HprogTocaRTA(p)\downarrow q_F')$ holds.
\end{itemize}

\section{Related work}
\label{sec:discussion}

Lindahl and Sagonas \cite{LindahlSagonas2006} have proposed success types out of the
consideration that standard type systems do no provide useful
guarantees for dynamically typed languages. They have defined and
implemented a widely used inference algorithm that is based on
constraint solving using fixpoint iteration.
Subsequently, they have considered the interplay of success types with
contracts and suggested a slicing-based algorithm for improved error
reporting \cite{SagonasSilvaTamarit2013}. 

Suter and coworkers \cite{DBLP:conf/sas/SuterKK11} introduce an
extension of Scala with statically checked contracts which are
provided by the programmer. These contracts may include recursively
defined functions and thus may be used for stating type
refinements. The contribution of the paper is the SMT-solver-based
contract checking procedure. In our work,
we are interested in inferring refinements in the form of success types.

Our approach draws a lot of inspiration from work on higher-order
model checking by Kobayashi and coworkers
\cite{Kobayashi2009,Kobayashi2009-popl,KobayashiIgarashi2013} as well
as by Ong and coworkers \cite{OngRamsey2011}. We are essentially applying
the procedure developed by Kobayashi \cite{Kobayashi2009-popl} to a
generalized automata model, the caRTA introduced in this paper. The
goal of the generalization is to obtain more precise analysis results;
the price is that our analysis is restricted to productive systems,
which may exclude certain programs from the analysis. We
need to gather further experience with our prototype implementation to
judge the impact of this restriction.

\section{Conclusion}
\label{sec:conclusion}

We propose a new approach for inferring success typings that uses context-aware
ranked tree automata (caRTS) and pattern-matching recursion schemes (PMRS) as models
to obtain more precise results than the algorithm of Lindahl and Sagonas.  We outline a
transformation from a first-order functional language to caRTS and PMRS and sketch an
extension of model checking by type inference for these models.

\section*{Acknowledgement}
This work has been partially supported by the German Research Foundation
(Deutsche Forschungsgemeinschaft, DFG)
within the Research Training Group 1103 (Embedded Microsystems).

\bibliographystyle{abbrv}
\bibliography{abbrv,papers,collections,theses,misc,books}

\end{document}